# The African Biophysics Landscape: A Provisional Status Report

Tjaart P.J. Krüger[a], B. Trevor Sewell[b], Lawrence Norris[c,d]

[a]Department of Physics, University of Pretoria, South Africa

[b]Department of Integrative Biomedical Sciences, University of Cape Town, South Africa

[c]African Physical Society

[d]African Light Source Foundation

**Abstract**

This is a provisional status report of biophysics activities in Africa. We start by highlighting the importance of biophysics research and development for every country's economy in the 21$^{st}$ century. Yet, the amount of biophysics activity in African countries varies between woefully little to nothing at all. We present a scope of biophysics research on the continent based on a pilot scientometrics study. We discuss a number of existing multinational programmes and infrastructure initiatives and propose a Pan African Professional Society for Biophysics. We emphasize the need for education, infrastructure and career development, and conclude with a list of suggested recommendations for expedited development of biophysics research on the continent.

## Introduction

Biophysics is an interdisciplinary field that applies the principles and methods of physics to understand how biological systems work. This research field brings the disciplines and concepts of physics and biology together in a unique manner and can be applied to all scales of biological organization, from the molecular level where quantum biology plays an important role to the level of organisms and populations of organisms. One can gain an understanding of how energy flows through vast complex systems to obtaining information and manipulating entities on the molecular and atomic levels. As such Biophysics is an extremely powerful scientific platform that can address many of the critical challenges that face humanity today and in the future. To undertake this work requires highly skilled individuals collaborating across the globe. These individuals, skills and collaborations need to be identified and fostered. Africa is a fundamental player in this endeavour with a vast, as yet relatively untapped resource of human capital that needs to be mobilized. This initiative is a critical element in this process and Africa is on the brink of a renaissance which must be encouraged and allowed to grow.

## The importance of Biophysics in Africa

Biophysics can have a tremendous impact on any country's economy. For example, by understanding crop yields and animal physiology at the molecular level, harnessing energy production and pollution remediation capabilities of plants and bacteria, and of course meeting all the human health challenges that cripple our lives. In fact, biophysics is a fundamental enabling science in medicine, agribusiness, and industrial biotechnology. [1] It is a wellspring of innovation and instrumentation for any country interested in developing a high-tech economy.

Biophysics has already revolutionised medical research and continues to do so, as evidenced by many Nobel Prizes. The previous century has evidenced great progress in treating diseases. In fact, biophysics helped to create powerful vaccines against infectious diseases. It described and controlled diseases of metabolism, such as diabetes. Biophysics provided both the tools and the understanding for treating the diseases of growth known as cancers. With the help of biophysics we are witnessing in the 21$^{st}$ century a rapid progress in the understanding of diseases at a fundamental level. [2] At the Biophysics Winter School of the 66$^{th}$ Annual Conference of the South African Institute of Physics on 1

July 2022, Martin Friede (the scientific officer responsible for vaccine delivery systems within the Initiative for Vaccine Research (IVR) at the World Health Organization in Geneva, Switzerland) stated that it is impossible to develop the next generation of vaccines without biophysics.

Biophysics underpins very large sections of the bio-economy and therefore a strong and diverse biophysics research and commercial sector is vital for the success of the African economy. There is a global emphasis on developing the bio-economy. The UK [3], EU [4], USA [5], and South Africa [6] have all formulated strategies to move away from the traditional industrial base and develop a strong bio-economy.

Biophysics is not limited to the bio-economy. Besides medical sciences and food security, other Grand Challenges that biophysics can address include environmental management (for example to combat pollution and climate change), energy security (to develop renewable energy sources), computing, and even security and telecommunications. Clearly, Biophysics drives numerous critical innovative developments.

The 21$^{st}$ century has been referred to as the *century of biology* [7], with the expectation that biotechnology will lead to some of the most significant innovations. One example is Quantum Biology, which is expected to play an important future role. In fact, many of the European Union's Key Enabling Technologies are based on quantum-mechanical phenomena and find examples in nature. Specifically, in the areas of photonics, nanotechnology, nanoelectronics, advanced materials, and photovoltaics there are examples of biological systems that can do what we would like to accomplish in man-made technologies.

## Scope of biophysics research in Africa

For an initial assessment of the scope of biophysics research in Africa, we conducted a pilot scientometric study. On the Web of Science database we selected all publications co-authored by at least one African researcher in journals that contain the keyword *\*Biophys\**, *\*Biomolecular\**, or *Biomaterials*. The first keyword represented mainly the journals Biophysical Journal, Biophysical Chemistry, European Biophysics Journal with Biophysics Letters, Biochemistry and Biophysics\*, and Biochimica et Biophysica Acta\*. The Journal of Biomolecular Structure and Dynamics was represented by the second keyword, while the journal Biomaterials was the main hit for the third keyword. The results are displayed in Figure 1, where the keyword *Biomaterials* was considered separately.

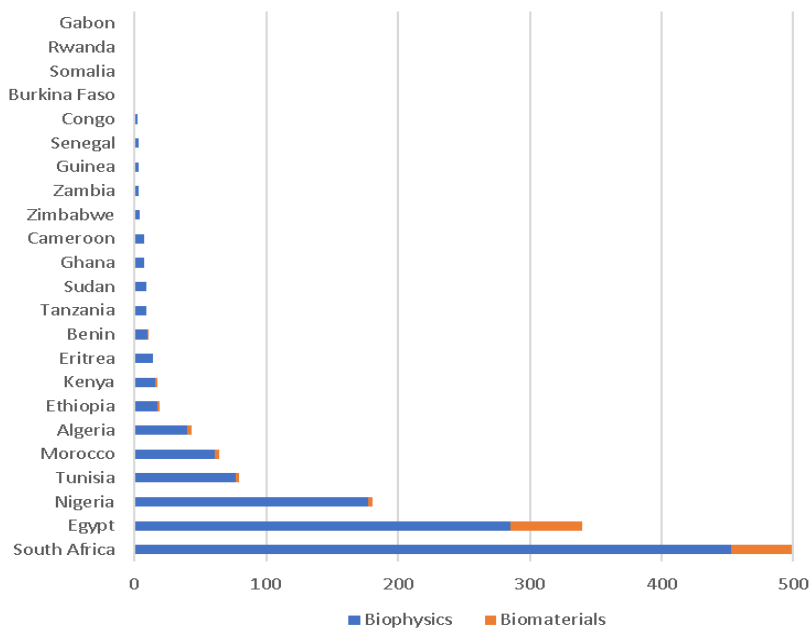

**Figure 1**: Representative WoS Biophysics Publications from Africa

Although this is by far not an exhaustive list of biophysics publications, we considered it representative for an initial assessment because these are well-known biophysics journals that publish predominantly biophysics research. We recognize that the relative output between countries may somewhat change in a more comprehensive study.

Our findings in Figure 1 indicate that South Africa is leading biophysics research on the continent. This is not surprising considering that this country, in general, produces most ISI scientific publications. By extension, the top few countries in Figure 1 are recognized as the most research-intensive countries on the continent. This study made us aware of a large number of active biophysics researchers at various institutions across the continent. At the other hand, we also recognize that the output of some countries such as Ethiopia, Kenya and Tanzania is possibly underestimated in Figure 1 because we are aware of a few institutions in these countries that are active in biophysics research.

We are eager to extend this pilot study because a comprehensive biophysics database would enable inclusive development of biophysics research on the continent. A comprehensive scientometrics study will also provide a significant amount of valuable information such as a breakdown of institutions and departments active in biophysics research, research topics within the broad domain of biophysics that are studied on the continent, better differentiation between research driven by African authors and those in which African researchers play a secondary role. An extended study may also include various other disciplines that have a strong overlap with biophysics, for example, biomathematics and bioinformatics. Our database of biomaterials research in Africa must also be extended. There is a comparatively large interest in biomaterials research in Africa, which is not reflected adequately in Figure 1.

The following are some of the challenges that must be recognized in a more comprehensive study. Biophysics is vast and interdisciplinary, with biophysics papers often published in different types of discipline-specific journals (such as physics, chemistry, biology, and engineering) and in various interdisciplinary journals (like physical chemistry, and chemical biology). The fraction of biophysics papers in those journals is often relatively small, indicating that it may be difficult to filter out biophysics-specific papers from databases without a significant amount of labour-intensive work.

We may also consider including non-ISI journals. Two examples are the *Egyptian Journal of Biophysics and Biomedical Engineering* and the *Nigerian Journal of Biophysics*, which we consider to be good initiatives for developing biophysics research in Egypt and Nigeria, even though the former journal publishes only about five papers per annum. We do, however, express caution about the Nigerian Journal of Biophysics, which shows characteristics of a possible predatory journal.

Through the Association of African Universities (AAU), the African Research Universities Alliance (ARUA), and the Network of African Science Academies (NASAC), we intend to field a survey of institutions to determine the extent of biophysics research at their institution as measured by publications in biophysics journals co-authored by faculty members, Honours, MSc, and PhD theses in the field, and graduate courses in biophysics, which may be definedbroadly.

**Multinational research programmes and meetings**

To put Africa on the global Biophysics maps, it is essential to establish multinational research programmes, consortia, and training events. Examples of existing successful research and training centres are the African Centre for Advanced Studies based at the University of Yaoundé I (ACAS, having computational biophysics as one of their research focus areas), the African Institute for Mathematical Sciences (AIMS, with mathematical biology as a research thread), the African Laser Centre (ALC, which funds many research projects on biophotonics), and the National Institute for Theoretical and Computational Sciences (NITheCS, with research programmes on bioinformatics and quantum biology).

Research in structural biology has seen notable development in recent years, in particular in South Africa, as a result of international collaborations. In general, these collaborations occurred between individuals and were stimulated by grants from international funding organizations including the Wellcome Trust and the Carnegie Corporation of New York. The Carnegie Corporation grant to the University of Cape Town and the University of the Western Cape was awarded for the establishment of a Master's programme in Structural Biology that was taught by internationally renowned experts. It also enabled the creation of rudimentary local infrastructure, gave rise to substantial research, and led to the continuing use of the EMBL beamline (BM14) at ESRF by South African scientists. More recently, the START (Synchrotron Techniques for African Research and Technology) programme funded by the UK Science and Technology Research Council funded research using the facilities at Diamond Light Source and eBIC (electron Bio-Imaging Centre) by nine PIs and thirteen bioscience postdoctoral research assistants over a period of three years [8].

The START programme led to the formation of a community of some 50 people with the common interest of using the facilities of the Diamond Light Source for protein structure determination. This, in turn, led to collaboration and cooperation. The challenge, following the end of the funding period, is to hold this community together. To this end, a meeting called "Legacy of START" was held in October 2022 to discuss the achievements of the START programme.

A further important development from the Legacy of START meeting is that a discussion with the South African National Research Foundation (NRF) on the development of a sustainable national programme in Structural Biology had been initiated. The NRF want to focus the discussion on the acquisition of a modern cryo-electron microscope and the supporting infrastructure required to make this viable.

Biostruct-Africa is another Structural Biology initiative. It was recently established as an independent organization to promote knowledge transfer and mentoring of Africa-based researchers using cutting-edge techniques in Structural Biology.

It is undeniable that meetings of international societies play an important role in bringing biophysicists together and thus enabling people to identify others with similar interests that may ultimately lead to collaborations and joint publications. In 2021, the African Light Source Foundation and the African Physical Society collaborated to put on the *Biophysics in Africa* online conference during Biophysical Society's Biophysics Week. There were ~40 presentations. The two organizations plan to reprise the effort in 2023, during the week of 20-24 March, again the Biophysical Society's Biophysics Week.

Three prominent international societies represent the interests of biophysicists, viz. the International Union for Pure and Applied Biophysics (IUPAB), the International Union of Pure and Applied Physics (IUPAP) through Commission C6: Biological Physics, and the Biophysical Society (BPS). While the first two require national membership, the BPS is an individual membership society. BPS have sponsored a thematic meeting on "Biophysics in the Understanding, Diagnosis, and Treatment of Infectious Diseases" in South Africa and IUPAB have sponsored a meeting on "Biophysics and Structural Biology at Synchrotrons" that was documented in a special issue of Biophysical Reviews [9].

IUPAB is keen to promote biophysics in Africa. They have recently held capacity building events in Kenya and Uganda and is currently constituting a Kenyan Society with currently 5 distinct and active Chapters established in 2022 around the country. Similarly, a Ugandan Biophysical Society (with one chapter to date) is being constituted. The National Committee for Physics in Egypt has a Biophysics subgroup, and recently approached IUPAB for active participation.

There are many other societies in which biophysicists participate and which assist in promoting overlapping interests. Notable among these is the International Union of Crystallography (IUCr) that assists in coordinating a substantial international network of Structural Biologists. Notably, after several years of development following the International Year of Crystallography in 2014, the African Crystallographic Association (AfCA) was founded at the annual conference of the African Light Source and African Physical Society (*vide infra*). AfCA will be convening its first stand-alone conference in Kenya in January 2023.

**Towards a Pan-African Professional Society for Biophysics**

Professional societies generally set the parameters in which people practice the profession. In some cases, professional societies control certification and accreditation processes, thus establishing who has the right to practice in the profession. Professional societies generally are the stewards of communication channels and archives of the profession. Thus, they are the mechanisms where people gain recognition in the profession. They set the standards for ethical professional practice. They are advocates for the profession in government and with various other venues. Professional societies are generally the mechanisms by which change is instituted in the profession. They provide forums for marketing, intellectual pollination, and career networking. They provide personal and professional services.

These functions are vital for biophysics in Africa. But with there being limited sustainable resources and given the multi-disciplinary nature of biophysics, it will be highly beneficial to cooperate with already existing disciplinary societies to establish a full-on, Pan-African biophysics professional society. We suggest forming an advisory council consisting of members of the following organizations: African Bioimaging Consortium, African Crystallographic Association, African Laser Centre, African Light Source Foundation, African Materials Research Society, African Mathematical Union, African Optics and Photonics Society, African Physical Society, Biostruct Africa, Federation of African Engineering Organizations, Federation of African Medical Physics Organizations, Federation of African Societies of Biochemistry and Molecular Biology, Federation of African Societies of Chemistry, IEEE Africa, and Society of Neuroscience in Africa.

With input and support from this advisory council, the activities of the Pan-African biophysics could include the following: (a) Convening meetings, schools, and workshops; (b) Operating a preprint server for African biophysics researchers to post the pre-published work; (c) Operating a repository of MSc and PhD theses in biophysics; (d) Operating a virtual journal that collects the published work of Africans in biophysics; (e) Publishing a newsletter about and for African biophysicists; and (f) Operating an online community of practice for African biophysicists, both on and off the continent. Inside this online community can be subgroup structure according to particular interests, e.g., biocomputing, biomaterials, biomechanics, biophotonics, cell physiology, clinical medical physics, genomics, industrial biophysics, mathematical biology, molecular biochemistry and pharmacology, physical biochemistry, plant and agricultural biophysics, quantum biology, and structural biology.

This council could immediately develop an implementation plan for the South African Bioeconomy Strategy in a Pan-African context as well as a Pan-African Structural Biology Initiative, and a strategy for exploiting Africa's biodiversity. The biophysics society can take on developing or being a continental adhering body of IUPAB as well as a Pan-African representative for IUPAP C6. Through an effort analogous to the ASFAP, the council can develop a 'Shaping' report specifically for biophysics and medical physics, so that the African governments can understand the intellectual and economic role of these two related but distinct fields [13].

**Infrastructure in Africa for Biophysics research**

Infrastructure is an indirect source of information about the quality and quantity of research activity. Although numerous state-of-the-art research facilities on other continents are accessible to African researchers, here we are focussing exclusively on research infrastructure on the African continent. A representative scientometric analysis of biomolecular research equipment at sub-Saharan universities for plants-based drug development found that there is a severe shortage of essential research equipment and that many of the existing research instruments are in a dysfunctional state [10]. The study focussed only on a sample of 25 universities where the research environment is relatively good. The situation at most other universities on the continent is even worse. The study concludes that "the cost of establishing comprehensive biomolecular research infrastructure in at least one university per sub-Saharan nation is negligible relative to their gross domestic products (GDPs). Thus, even with the current economic resources, sub-Saharan African countries would upgrade biomolecular research capabilities in their leading universities without disrupting other economic priorities." In other words, sub-Saharan countries should see no financial barrier against buying essential research equipment. What is needed is political impetus, which may start by influencing policymakers and making them aware of the essential need for basic biophysics infrastructure.

The African Laser Centre (ALC) is an example of an infrastructure programme of benefit to the whole continent. The programme provides funding for accessing and maintaining laser research equipment, for laser-based research collaborations on the continent, and for postgraduate student bursaries. the

The programme's purpose is to support laser-based research and technological development in the whole of Africa provided that access to infrastructure, support, and funding proceeds via a South-African based researcher. Noteworthy is that Biophotonics is a growing research domain of this programme.

The Centre for High-Performance Computing (CHPC) based in Cape Town is an example of a South African-based infrastructure programme that makes its facilities available to various other African countries without the condition of an active research collaboration with a South-African researcher. Researchers of the 16 SADC Members States and the 8 African SKA Partner Countries may use the CHPC for any research domain. Researchers from remaining African countries have access to the CHPC through collaboration with South African researchers. Currently, only a small number of projects for which the CHPC is used have a biophysics nature. In the near future, we would like to see a significant growth in computational biophysics enabled by the CHPC. Computational biophysics is also undoubtedly enabled by the African Supercomputing Center (ASCC) in Morocco, and the Laboratory of Modelling and Simulation in Engineering, Biomimetics and Prototypes at the University of Yaoundé I in Cameroon.

In South Africa, the NRF provides for the National Equipment Programme (NEP) that co-funds equipment at academic institutions. This has resulted in widespread placement of equipment that is of interest to biophysicists including X-ray diffractometers, electron microscopes, mass spectrometers, confocal microscopes, a single-molecule spectroscopy facility, and a variety of other important pieces of support equipment.

Recently, the Chan Zuckerberg Initiative has established the Africa Microscopy Initiative that houses a fleet of advanced, commercial optical microscopes [11]. These instruments are tailored for use in the life sciences and provide high spatial and temporal resolution to study a variety of biology. Equipped with ancillary infrastructure that includes cell culture and molecular biology equipment, the Imaging Centre offers advanced microscopy instruments on an open-access basis to scientists working in Africa. Coupled to the centre is a training programme that will support African Scientists in the achievement of their research goals without any cost to the researcher and the African Bioimaging Consortium (ABIC), a network of life science researchers throughout Africa with an interest in and use for microscopy. This community-driven initiative aims to unite the African microscopy community and support the development of bioimaging across the continent.

The Benin government established the X-TechLab in Seme City research park in collaboration with IUCr. X-TechLab is an experimental platform dedicated to the use of X-ray techniques for scientific and technological research on development issues specific to the African continent. It has held training courses in X-ray crystallography, and supported research in small-molecule crystallography, and structural enzymology [15,16].

In spite of these encouraging developments there are gaping holes in the infrastructure available to biophysicists – the most notable being the complete absence of modern NMR instruments and cryo-electron microscopes on the continent. The fragility of the situation in cryo-electron microscopy compromises the ability of African scientists to use this technique at international infrastructures as the research samples cannot be validated and thus access will not be granted.

Less well known is the desperate shortage of modern equipment for the preparation of proteins. As a consequence, many researchers resort to leaving the preparation of material to international collaborators, thus weakening the position of the African scientists. This situation results in generations of students that are unable to reliably produce the basic material required for further research.

It is clearly impossible for the continent to house all the equipment necessary for modern biophysics and provide the expertise necessary to exploit and maintain it. This is also true for most European countries. The European solution is to form consortia called ERICs (European Research Infrastructure Consortium), which facilitate the establishment and operation of research infrastructures with European interest on a not-for-profit basis. The ERIC of interest to Structural Biologists is Instruct, which provides access to 11 centres and 27 facilities in 14 countries for scientists from member countries. The resources can also be accessed by scientists from non-member countries. South African scientists have exploited the resources through an MoU with the University of Cape Town.

**Career Development**

The world has seen a substantial growth in biophysics and structural biology over the last half century. An evident example for this is the growth in the number of deposits in the protein data bank, which has now reached nearly 200,000. Macromolecular structures are the enablers of studies in molecular biophysics and the numbers are thus a good indicator of activity. Furthermore, the structures are widely used by industries concerned with drug and agrochemical development. The African contribution to this huge international effort is negligible and has come from only two countries: South Africa and Egypt [12]. Indeed, very few African biophysicists and structural biologists have found work on the continent and most have built their careers abroad. It is self-evident that the growth of these disciplines in Africa depends on the retention and development of qualified people.

**Recommendations**

We offer the following list of suggestions for accelerated development of biophysics research on the continent:

1. **Establishment of a Pan-African Professional Society for Biophysics**, as discussed above.
2. **Establishment of sustainable academic infrastructure**: Considerable investment is required to extend biophysics infrastructure on the continent to promote state-of-the-art home-based research. Additional funding is required to maintain existing infrastructure and to replace initial appointees following their resignation or retirement.
3. **Scientometrics/bibliometrics study**: Extension of the pilot scientometrics study to create a comprehensive database of biophysicists in Africa and a catalogue of biophysics research, training, and education happening on the continent.
4. **Tapping into support from international societies**: IUPAB has expressed their desire to establish more biophysics activity on the African continent, for example by supporting and co-organizing workshops or even congresses in the future. At present only South Africa maintains its membership of the IUPAB through the South African Institute of Physics (SAIP). IUPAB supplies up to EUR 10,000 start-up grants for workshops and conferences. They have 61 affiliated societies and over 20,000 members.
5. **Biophysics schools, workshops and conferences**: Besides the activities already taking place, for example an annual Biophysics School in Tanzania, Biophysics Schools organized through SAIP, and several Schools and Workshops in Structural Biophysics, we will encourage more activities. One of them is an annual Biophysics in Africa online conference during Biophysics Week, which was started in 2021.
6. **African research libraries**: We want to encourage African research libraries to have a full complement of biophysics titles in their serial subscriptions.
7. **Biophysics careers**: We would like to create a document listing potential jobs when studying biophysics, emphasizing that biophysics not only contributes to the knowledge industry but also to important economic development of every country.
8. **Virtual department**: Accessibility of biophysics education can be greatly enhanced by creating a virtual African biophysics department where a network of people can give online courses and engage with students online.
9. **Requests for political incentives**: African governments should develop multi-departmental initiatives to support the work of African biophysicists. They should incentivize our universities to build infrastructure in all the fields that support biophysics, including chemistry and biochemistry labs, computing, as well as equipment in spectroscopy, electron microscopy and crystallography. They should implement policies that encourage our industries to invest in the bioeconomy strategy.
10. **Encouragement**: The development of biophysics research on the African continent has been slow over the past few decades. A major reason is that most African people interested in biophysics study abroad and remain abroad. Most of those who returned to their home countries have remained in biophysics for short periods of time. There is therefore a great need of nurturing and encouraging existing and aspiring biophysicists.

## Acknowledgments

The authors would like to thank Raymond Sparrow for fruitful discussions and feedback.
## References

1. https://www.dailymaverick.co.za/article/2018-04-04-op-ed-south-africa-needs-more-investment-in-biophysics-and-structural-biology/
2. https://www.up.ac.za/media/shared/107/Research/Biophysics/saip-biophysics-brochure.zp79597.pdf
3. Growing the Bioeconomy. Improving lives and strengthening our economy: A national bioeconomy strategy to 2030. HM Government, UK (2018)
4. A sustainable bioeconomy for Europe: strengthening the connection between economy, society and the environment. *Updated Bioeconomy Strategy*. EC (2018)
5. National Bioeconomy Blueprint. The White House, Washington, USA (2012)
6. https://www.gov.za/sites/default/files/gcis_document/201409/bioeconomy-strategya.pdf
7. C. Venter and D. Cohen, The Century of Biology, *New Perspectives Quarterly* 21(4): 73-77 (2004), DOI: 10.1111/j.1540-5842.2004.00701.x
8. C. Nicklin, R. Stredwick and B.T. Sewell, Synchrotron Techniques for African Research and Technology: A Step-Change in Structural Biology and Energy Materials, *Synchrotron Radiation News* 35(1): 14-19 (2022), DOI: 10.1080/08940886.2022.2043684
9. B.T. Sewell, The workshop on "Biophysics and Structural Biology at Synchrotrons" presented at the University of Cape Town from 16–24 January 2019. *Biophys. Rev.* 11(4): 491–493 (2019). https://doi.org/10.1007/s12551-019-00575-6
10. P.F. Kamba *et al.* Slow translation of Tropical Africa's wealth in medicinal plants into the clinic: Current biomolecular infrastructural capacity and gaps in sub-Saharan universities, *Sci. Res. Essays* 11(17), pp. 174-186 (2016) DOI: https://doi.org/10.5897/SRE2016.6435
11. Africa Microscopy Initiative: https://www.microscopy.africa/; https://www.news.uct.ac.za/article/-2021-11-15-grant-to-boost-bioimaging-in-africa
12. S. Burley *et al*. Protein Data Bank: A Comprehensive Review of 3D Structure Holdings and Worldwide Utilization by Researchers, Educators, and Students. *Biomolecules* 12:1425 (2022). DOI: 10.3390/biom12101425
13. Shaping the Future of Physics in South Africa, Report of the International Panel Appointed by the Department of Science and Technology, National Research Foundation, and South African Institute of Physics, April 2004, https://www.saip.org.za/images/stories/documents/IP%20Report%20-%20Executive%20Summary.pdf
14. S. Baammi, R. Daoud and A. El Allali. Assessing the effect of a series of mutations on the dynamic behavior of phosphite dehydrogenase using molecular docking, molecular dynamics and quantum mechanics/molecular mechanics simulations. *J. Biomol. Struct. Dyn.* 20:1-13 (2022). DOI: 10.1080/07391102.2022.2064912
15. M. Agbahoungbata *et al.* X-TechLab training sessions in Benin: towards borderless science education. *Acta Cryst.* A77, C606 (2021).
16. F.H. Agnimonhan *et al.* Crystal structure of a new phenyl (morpholino) methanethione derivative: 4-[(morpholin-4-yl) carbothioyl] benzoic acid. *Acta Crystallogr. E Crystallogr. Commun.* 76(4): 581-584 (2020). DOI: 10.1107/S2056989020003977